%% file: PCA_v1.tex
\documentclass{aa}  

\usepackage{graphicx}
\usepackage{txfonts}
\usepackage{hyperref}

\usepackage{xcolor}
\usepackage{soul}

\begin{document}

   \title{Improving mid-infrared thermal background subtraction with Principal Component Analysis}


     \author{
        H.Rousseau\inst{\ref{lbti},\ref{ago}},
        S.Ertel\inst{\ref{lbti},\ref{steward}},
        D.Defrère\inst{\ref{ku}},
        V.Faramaz\inst{\ref{lbti},\ref{steward}}
        \and
        K.Wagner\inst{\ref{lbti},\ref{steward}}
    }

    \institute{
        Large Binocular Telescope Observatory, University of Arizona, 933 N Cherry Ave., Tucson, AZ 85721-0065, USA\\
        \email{hrousseau@lbti.org}\label{lbti}
        \and
        AGO Department, University of Liège, Allée du 6 août, 19C B-4000 Liège 1,Belgium \label{ago}
        \and
        Institute of Astronomy, KU Leuven, Celestijnenlaan 200D, 3001 Leuven, Belgium \label{ku}
        \and
        Department of Astronomy and Steward Observatory, University of Arizona, 933 N Cherry Ave., Tucson, AZ 85721-0065, USA \label{steward}
    }

   \date{}

   \titlerunning{PCA for mid-IR background subtraction}

   \authorrunning{Rousseau et al.}

 
  \abstract
   { Ground-based large-aperture telescopes, interferometers, and future Extremely Large Telescopes equipped with adaptive-optics systems provide angular resolution and high-contrast performance that are superior to space-based telescopes at thermal-infrared wavelengths. Their sensitivity, however, is critically limited by the high thermal background inherent to ground-based observations in this wavelength regime.}
   {We aim to improve the subtraction quality of the thermal-infrared background from ground-based observations, using Principal Component Analysis (PCA).}
   {We use data obtained with the Nulling-Optimized Mid-Infrared Camera on the Large Binocular Telescope Interferometer as a proxy for general high-sensitivity, AO-assisted ground-based data. We apply both a classical background subtraction -- using the mean of dedicated background observations -- and a new background subtraction based on a PCA of the background observations. We compare the performances of these two methods in both high-contrast imaging and aperture photometry.}
   {Compared to the classical background subtraction approach, PCA background subtraction delivers up to two times better contrasts down to the diffraction limit of the LBT's primary aperture (i.e., 350 mas in N band), that is, in the case of high-contrast imaging. Improvement factor between two and three are obtained over the mean background retrieval within the diffraction limit in the case of aperture photometry.}
   {PCA background subtraction significantly improves the sensitivity of ground-based thermal-infrared imaging observations. When applied to LBTI's nulling interferometry data, we expect the method to improve the sensitivity by a similar factor 2-3. This study paves the way to maximising the potential of future infrared ground-based instruments and facilities, such as the future 30m-class telescopes.}


   \maketitle


\input{Sections/intro}




\input{Sections/data}






\input{Sections/results}


\input{Sections/discussion}


\input{Sections/conclusion}




\begin{acknowledgements}
HR, SE, and VF are supported by the National Aeronautics and Space Administration through the Exoplanet Research Program (Grant No. 80NSSC21K0394).
DD acknowledges the support from the European Research Council (ERC) under the European Union's Horizon 2020 research and innovation program (grant agreement CoG - 866070). 
\end{acknowledgements}

%
%

\bibliographystyle{aa.bst}
\bibliography{PCA_v1.bib}

\end{document}

%% file: Sections/intro.tex
\section{Introduction}

Thermal infrared wavelengths ($3-13\,\mu$m)  make up an observational regime that has become essential to unravel and study a wide variety of astronomical objects, ranging from Active Galactic Nuclei to exoplanets and Solar system bodies. 
{At these wavelengths, the angular resolution reached by spaced-based telescopes -- measured as $\lambda/D$, where $\lambda$ is the wavelength, and $D$ is the diameter of the primary mirror -- is lower than existing and future large, ground-based telescopes. 
For example, the {\it Spitzer Space Telescope} had a primary diameter of $D = 85\,$cm (\citealt{Spitzer}), and the {\it James Webb Space Telescope} {\it JWST}, \citealt{JWST}) has $D = 6.5\,$m,  while current large ground-based telescopes such as the Very Large Telescope (VLT, \citealt{VLT}) and the Large Binocular Telescope (LBT, \citealt{LBT}) have respectively $D = 8.2\,$m and $D = 8.4\,$m. Furthermore the Large Binocular Telescope Interferometer (LBTI) can, thanks to optical interferometry, reaches the angular resolution of a 22.65 m telescope (\citealt{Hill_2006_SPIE}). As for future extremely large telescopes (ELTs) their sizes range from $D = 25.4\,$m, (Giant Magellan Telescope, GMT, \citealt{GMT}) to $39.3\,$m, (European Extremely Large Telescope, E-ELT, \citealt{ELT}).  
Their sensitivity is, however, limited by the high thermal background due to photon noise, and to the imperfect removal of background structures from both the sky and warm telescope optics (\citealt{defrere_2016}, \citealt{ertel_hosts_2020}). 
Therefore, in order to unlock the full potential of existing and future large, ground-based telescopes operating at thermal infrared wavelengths, it is paramount to develop methods that effectively remove spatially and temporally variable background structures. 
In particular, this is essential to the field of exoplanetology, as it now enters the characterization era. 
The new generation of telescopes such as the JWST and 30m-class telescopes such as the E-ELT, the GMT and the Thirty Meter Telescope (TMT, \citealt{TMT}) will focus on identifying the components of exoplanets' atmospheres and surface conditions (temperature, pressure, composition, ...). 
In order to characterize Earth-like exoplanets, orbiting within their host star's habitable zone, direct imaging will need to address three main challenges: the sensitivity, the contrast ($10^{-7}$ in the $N$' band for an Earth analogue, \citealt{2017Msngr.169...16K}, \citealt{2023AJ....165..133W}) and the small inner working angles (10 milli-arcseconds to 1 arcsecond for a planet in the habitable zone at 10 pc depending on the host star's luminosity, \citealt{2017Msngr.169...16K}, \citealt{2023AJ....165..133W}).

The N-band (8-13um) is especially relevant here because temperate, habitable-zone exoplanets strongly emit in this wavelength range and their contrast to their host stars is particularly favorable (\citealt{2017Msngr.169...16K}).  This wavelength range has only recently been opened up to adaptive-optics, high-contrast imaging by the availability of adaptive secondary mirrors (e.g., \citealt{AO}). It is, however, especially challenging to observe in from the ground due to the high background. Thus an effective removal of this background is particularly important to reach high sensitivity and exploit the full capacities of the instruments.

\cite{hunziker} have explored a method based on the Principal-Component-Analysis (PCA) to better handle the thermal background in the $L$ and $M$ bands for high-contrast imaging observations. PCA is a statistical method which allows one to reconstruct the data as a basis of eigenvectors corresponding to the level of variance in the data. It thus determines the dominant features of this set of data. The number of principal components used in the analysis determines the strength of the features that can be considered, from the most dominant to the less significant ones. Thus, it determines the level of detail the PCA analysis will be able to reconstruct. This method can significantly improve the background subtraction compared to the more common method of subtracting the (selective) mean image from dedicated background exposure \citep{hunziker}. In this paper, we expand this study to the $N$’ band and apply it to both high-contrast imaging and the broader field of aperture photometry. 

In Section~\ref{sec:data}, we present the data and their primary technique of reduction. We present our findings in Section~\ref{sec:results}, discuss these results in Section~\ref{sec:discu}, and summarize our conclusions in Section~\ref{sec:conclu}.

%% file: Sections/data.tex
\section{Raw data and data reduction}\label{sec:data}

\begin{figure}[ht]
    \centering
    \includegraphics[scale = 0.4, trim = 70 110 0 0 ]{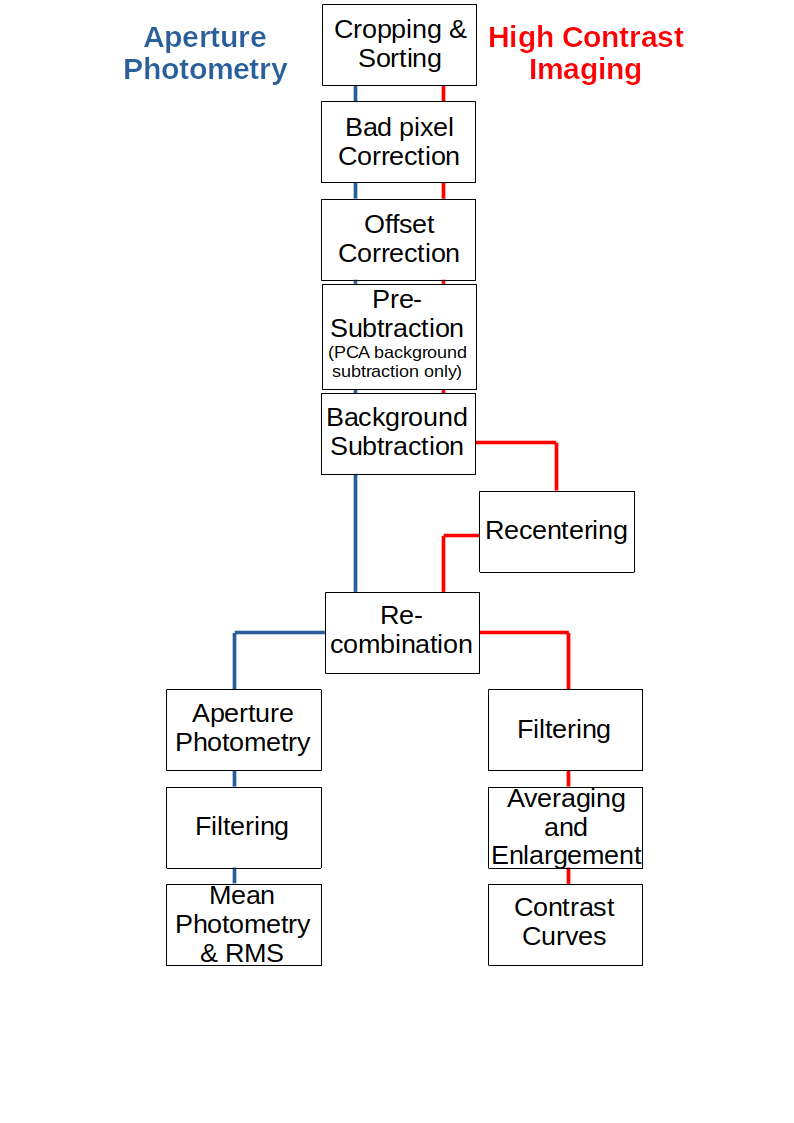}
    \caption{Summary of the data reduction processing steps for Aperture Photometry and High-contrast Imaging.}
    \label{fig:steps}
\end{figure}

In this section, we present the LBTI data we use for our analysis and the different steps of our data reduction. Figure \ref{fig:steps} shows a flow chart of the performed individual steps. These steps are described in more details below. 

\subsection{HOSTS/LBTI data}

The data used in this work were obtained during the commissioning time (February 2014 - November 2015) of the Hunt for Observable Signature of Terrestrial planetary Systems  (HOSTS, exozodiacal dust survey, \citealt{ertel_hosts_2018}, \citealt{ertel_hosts_2020}, September 2016 - May 2018). The observations were performed with the Nulling-Optimized Mid-Infrared Camera (NOMIC, \citealt{2014SPIE.9147E..1OH}) on the Large Binocular Telescope Interferometer (LBTI, \citealt{hinz_overview_2016}, \citealt{2020SPIE11446E..07E}) in the $N'$ filter ($\lambda_c = 11.11\,\mu$m, $\Delta\lambda = 2.6\,\mu$m). These data were taken in the context of the HOSTS survey with the nulling-interferometry mode of the LBTI. They can be used as a proxy for general, AO-assisted N-band imaging data for both high-contrast imaging and aperture photometry. This method provides a significantly improved contrast close to a star compared to regular adaptive optics (AO) imaging as it interferometrically suppresses the star light.  It thus allows us to detect fainter companions or circumstellar disks at smaller inner working angles than plain imaging. Furthermore, unlike most interferometers, the LBTI behaves like an imaging instrument with pupil stabilisation. The sky thus rotates across the detector during the observations and this field rotation can be exploited for high-contrast imaging (HCI) using techniques such as angular differential imaging (ADI).

The data used in this work were obtained following the HOSTS observing strategy and were detailed by  \cite{defrere_2016}. The observations of the science targets were separated by calibrator observations, using several calibrators for each scientific target. Each observation of a star consists of three parts: (1) nulling observations interfering the star light by overlapping the images from the two apertures in the pupil plane and offsetting between two detector positions (nodding) for background observations typically every 5-10 minutes (2) a photometric observation placing the star images from the two apertures next to each other on the detector, and (3) a background observation for the latter where the star images from the two apertures are moved off the detector. 

\begin{figure}[ht]
    \centering
    \includegraphics[scale = 0.37, trim = 50 0 0 0 ]{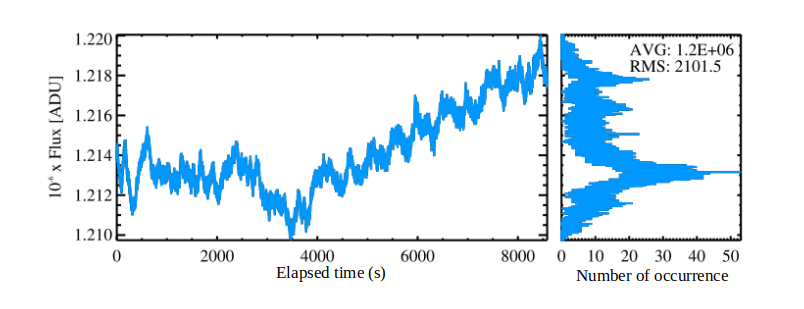}
    \caption{Example of on-sky raw thermal background measurements obtained in the $N'$ band with the telescope pointing at an empty region of the sky and covering approximately 15 degrees of elevation change during the whole duration of the sequence.The left panel shows the flux integrated over a photometric aperture of 8 pixels in radius while the right panel shows the corresponding distribution. The complete figure and a more detailed analysis of the performances of the nulling data-reduction performances for the HOSTS survey can be found in \cite{defrere_2016}.}
    \label{fig:timescale}
\end{figure}

A major sensitivity limitation of the HOSTS data comes from the strong thermal background in the $N'$ band. The timescale of the background variability can ranges from 0.1 seconds to several hours. Figure \ref{fig:timescale} shows an example of on-sky raw thermal background measurements with the LBTI. Comparing our timescale of 5-10 minutes for background calibration, we see that the background varies a lot within this timescale. The background subtraction effectively removes the background variation with a timescale larger than our background calibration timescale but will struggle with shorter timescales. Suppressing the variations at shorter timescales than the nodding period is a main goal of the present work. 

\subsection{Data sets}

\paragraph{\bf $\beta$ Leo dataset}\label{sec:data_HCI}

For high-contrast imaging, we used a dataset of $\beta$ Leo, which is composed of two sub-datasets. These two sub-datasets were taken the same night, on the UT 2015 February 8, along with their calibrators: HD104979, HD108381 and HD109749. These sub-datasets are respectively composed of 8000 and 8800 frames with 60ms exposure time with an offset every 1000 frames (except one group of 1800). The parallactic angle ranges of the two subdatasets are respectively from 41.11° to 45.02° and from 53.67° to 57.41° and the smallest distance from the star center to the edge of the usable field of view corresponds to a distance of 0.6 arcsec.

\paragraph{\bf Background-only dataset}\label{sec:data_ApPhot}

For aperture photometry, the dataset was taken on UT 2015 November 11, is a background-only dataset, composed of 24000 frames without any sky offsets. For this dataset, we recreated artificial groups using the 1000 frames per group model from the first dataset of $\beta$ Leo. We thus alternate groups of "source" and "background" exposures as if a star was present.

\subsection{General data reduction steps}\label{sec:image_prep}

The nodding sequence for our observations is composed of a top and a bottom position. We only keep the two left quadrants as the right part of the detector is not used. We will refer to the two left quadrants as top and bottom images. The final size of these sub-frames is 123x123 pixels (pixel size: 17.9 milli-arcseconds). We then sort the images so the top and bottom images can be treated separately, and then sort them per groups. For $\beta$ Leo, we thus have 8 groups for each part for both datasets. The groups of each part of the image alternate between ones with the star in them (source exposure) and without star (background exposures) due to the nodding between groups. 
After this first step we apply a bad pixel correction and a subtraction of the mean flux of each image from each individual pixel in that image (Offset Correction in Figure \ref{fig:steps}).

\subsubsection{Mean Background Subtraction}\label{sec:meanBS}

In the case of the mean background subtraction approach, we compute the mean of a sequence of dedicated background exposures and then subtract it from every on-source exposure. The high-contrast analysis, to search for circumstellar emission, is then performed without further treatment of the background. 
The aperture photometry analysis is performed using a background annulus to estimate the background under the photometric aperture. The photometry in this annulus is then subtracted from the photometry in the region of interest. The resulting photometry is then used for further analysis. This approach is extensively described in \citet{defrere_2016}.

\subsubsection{ PCA Background Subtraction}\label{sec:PCABS}

\paragraph{\bf Pre-subtraction}\label{sec:pre_subtraction}

Before performing PCA background subtraction, it is useful to remove static spacial background structures using a pre-subtraction. This is beneficial because we must use a mask to prevent over-subtraction from the star when computing the actual PCA for background subtraction. However, the mask strongly reduces the information available for principal-component building. The pre-subtraction effectively reduces this loss of information and helps build an optimal background subtraction. We do this pre-subtraction by subtracting a constant background frame from each individual image of a group. This image is derived from the background exposures only, without any mask. This step is similar to the mean background subtraction. In this case, we compute the PCA correction for the background exposures in the library, average the corrections images obtained and subtract this mean image from every on-source exposures.

\paragraph{\bf  Background subtraction}\label{sec:PCABSS}

The PCA approach is commonly used in high-contrast imaging to remove the stellar Point Spread Function (PSF, \citealt{amara2012}, \citealt{2012ApJ...755L..28S}, \citealt{amara2015}). Here, we use it to better estimate and remove the sky background, similar to the approach used by \cite{hunziker}. We thus perform PCA on background exposures. This determines the eigenvectors (or principal components), onto which the on-source images are projected. In order to avoid source over-subtraction, a mask on the star is introduced during this step (see below). With this technique, an optimal correction is computed for, and subtracted from, each on-source science image. This technique allows for an additional background removal to the classical subtraction for high-contrast analysis. In the case of aperture photometry, it replaces the background annulus used in the mean background subtraction. PCA is indeed capable of reconstructing the background at the position of  the star, which constitutes a more sophisticated estimate than the background annulus. 

\begin{figure*}[htbp]
    \centering
    \includegraphics[width=1\linewidth, trim = 0 70 0 30 ]{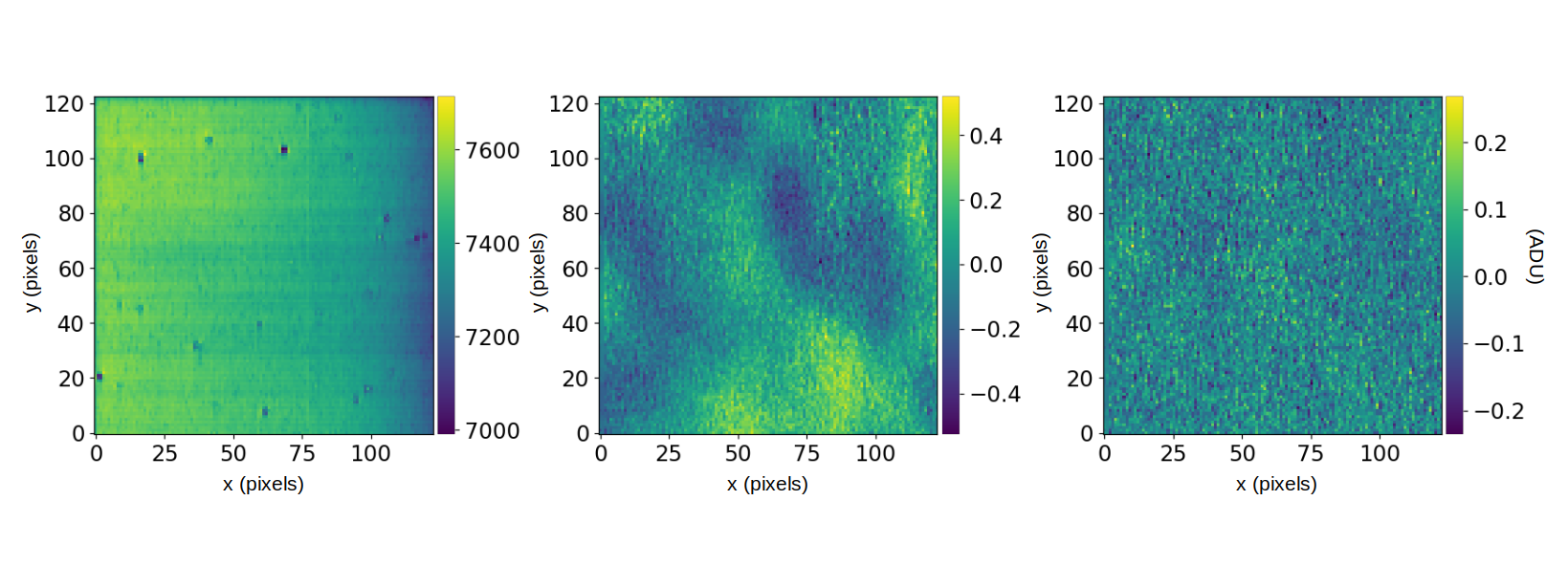}
    \caption{Raw image (left), the mean image of the dataset with a mean background subtraction (middle) and the mean image of the dataset with a PCA background subtraction (right). For the PCA background subtraction, we used a mask of 32 pixels in radius to mask the central region where the star would be located if any was present in the data.}
    \label{ImVis}
\end{figure*}

\paragraph{\bf Masking the star}\label{sec:masking_star}

A problem related to PCA background subtraction is over-subtraction since the astronomical source (in our case a star) is present to correct. To avoid over-subtraction, we thus mask the region of interest. We do not introduce a mask for mean background subtraction since the process never involves any information from the source exposure. The mask is necessary for PCA, but limits the region of the source images that can be used for projecting the principal components. In addition, if PCA computes the principal components on unmasked data, it makes use of masked data to determine the PC weights used in the correction. In a perfect case, with no astronomical source however, the coefficients too would be determined using the unmasked data. One cannot guarantee that those coefficients obtained with the masked or unmasked data would be the same. The introduction of a mask therefore breaks the rule of orthogonality in PCA. This introduces an error in our correction and is a limiting factor on the PCA performance presented in this study. 

Furthermore, mask sizes are ranging from 9 to 32 pixels in radius in our analysis, depending on the use case. While we can expect the smallest masks to have a very limited impact on the coefficients, the larger the error introduced when applying the background subtraction. We thus expect our data with a mask of 9 pixels in radius to retrieve  the coefficients with a better reliability, and thus to provide a correction of high quality, compared to those with a mask of 32 pixels in radius. To check that the use of the mask does not significantly impair PCA-background subtraction, we have applied our method to the background-only dataset. The results show that the correction is not optimal anymore, even on the smallest mask size. However, the effect is generally relatively small and will not impact significantly the results presented in this study.

At the end of our background subtraction, we recombine both top images and bottom images together to obtain a single dataset. This dataset is thus composed of the sub-frames selected in each original frame. This step is performed either just after background subtraction (aperture photometry) or after frame recentering the star (HCI).

\subsection{Application to high-contrast imaging}\label{sec:appli_HCI}

For HCI the $\beta$ Leo datasets need to be prepared for PSF subtraction. Thus, before recombining the datasets, all frames are re-centered so the star is at the center of each frame. The next step, after the dataset recombination consists in discarding the bad frames with high null-leak. This step is optional as unfiltered datasets might provide better performances depending on the configuration. After this filtering step, in order to save computational time, we average the frames by groups of 100 images. 

\subsection{Application to aperture photometry}\label{sec:appli_ApPhot}

For aperture photometry, we use the background-only data set. As we would have for a data set containing an astronomical source, we used a mask during PCA background subtraction, and a background annulus for mean background subtraction. Since the source can be extended, we explore a range of circular photometric aperture sizes and matching mask and background annulus sizes. We explore two cases.  In the first case, the photometric aperture matches the mask, i.e., the whole source flux is measured. This may be applied for both resolved and unresolved sources. For the second case, we still mask the entire (expected) source emission, but use smaller apertures of varying size as could be used to determine the radial distribution of extended emission or a curve-of-growth aperture photometry. This case, in particular applies to the HOSTS data of exozodiacal dust observations. For this last case, we use three different aperture sizes (8 and 13 pixels in radius, and a conservative aperture which cover the whole emission) as used during the HOSTS survey analysis ( \citealt{ertel_hosts_2020}, \citealt{defrere_2021}). However, to prevent the background correction to take any extended emission into account, the mask must cover the whole emission, even for smaller aperture sizes.

%% file: Sections/results.tex
\section{Results}\label{sec:results}

A first interesting result to display is a visual comparison of the image quality after mean or PCA background subtraction using a mask of 32 pixels in radius. In Figure \ref{ImVis}, we display a raw image, along with the mean image of the background-only dataset when the background subtraction is performed with the mean method and the mean image of the same dataset when its background subtraction is performed with PCA. The structures present in the mean image of the dataset when the background subtraction is performed with the mean method, almost completely disappear when the background subtraction is performed with PCA. We can also see that despite the use of a mask for the PCA background subtraction, the image is much cleaner even under the mask. This first result thus suggests that both the photometric measurement and the high contrast analysis will benefit from the PCA background subtraction.

\subsection{Contrast curves analysis: ADI, RDI, without PSF subtraction}\label{sec:CC}

To estimate the improvement obtained with PCA background subtraction we compare the results obtained from mean and PCA background subtraction using the contrast curves obtained with three different processing techniques: ADI, RDI but also without PSF subtraction and simply proceeding to a derotation of the frames. For ADI and RDI, the PSF is subtracted with a full-frame PCA algorithm from the Vortex Image Processing (VIP, \citealt{gonzalez_vip_2017}) library. It is important to distinguish PCA background subtraction and full-frame PCA PSF subtraction. The PCA for background subtraction and the PCA for PSF subtraction do not use the same principal components number. Furthermore, the mask is only used for background subtraction, and not for PSF subtraction.

After PSF removal, the contrast curves are computed using a 5 sigma threshold and the throughput is obtained through fake companion injection. The fake companion injection is performed after background subtraction. Therefore, the throughput does not take into account possible subtraction of a potential companion during the background subtraction step. However, since the background correction is computed on off-sources images, no signal from a potential companion is present in the library use to build the principal components. Hence, no planet signal is included in the principal components. Thus, self-subtraction can be ruled out. Projection of the PCs onto the science images may still result in an over-subtraction of the background around the potential companion if its presence can amplify some background signal in the PCs.  However, this effect is expected to be negligible since a the companion is typically be very faint and the background dominates the projection of the small number of PCs. This conclusion is amplified by the fact that the companion does not stay at the same position, with the parallactic angle range of the dataset. Given these considerations, we consider that the companion would not be self-subtracted or over-subtracted during the background subtraction step.

\begin{figure}[t]
    \centering
    \includegraphics[scale = 0.24, trim = 40 0 0 0]{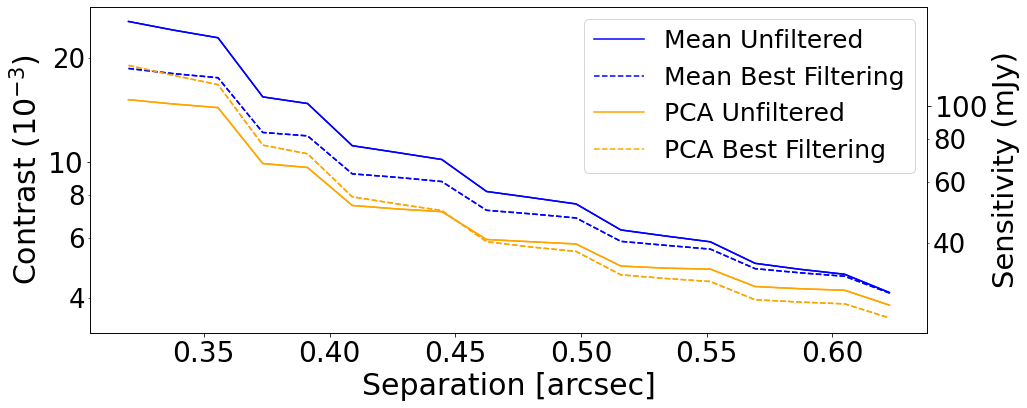}
    \includegraphics[scale = 0.24, trim = 30 0 0 0]{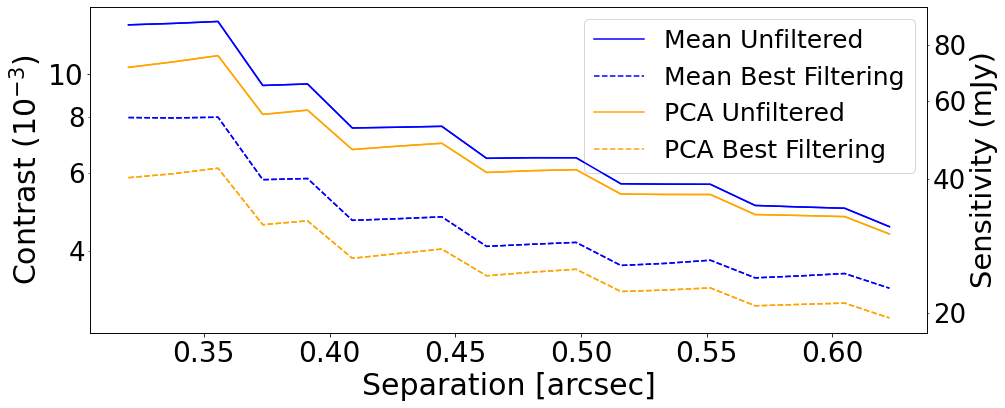}
    \includegraphics[scale = 0.24, trim = 30 0 0 0]{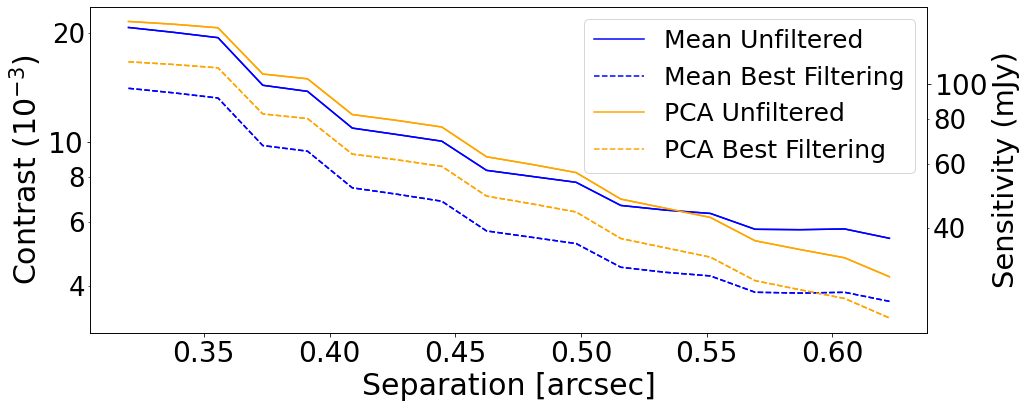}
    \caption{Contrast surves obtained with ADI (top panel), RDI (middle panel) and without PSF subtraction (bottom panel) without (straight lines) and with optimal filtering (dashed lines) for mean background subtraction (blue) and PCA background subtraction (orange). A pre-subtraction with mean has been applied on the PCA background subtracted cube before we applied the 9-pixel mask on the star.}
    \label{CC}
\end{figure}

For the ADI method, the PSF has been removed using the full-frame PCA algorithm for PSF subtraction provided by the VIP library. These results are presented in the top panel of Figure \ref{CC}. The solid lines represent the best contrast curves obtained among the different pre-subtractions and principal component numbers used for the background subtraction. We used for both solid and dashed lines and for both mean and PCA background subtractions 10 principal components for the PSF subtraction as the convergence of the contrast curves is reached at this point for all the configurations compared here. The dashed lines represent the best contrast curves obtained among the different aforementioned configurations but also among the different filtering levels. For the purpose of this analysis we tested four filtering levels: 80, 60, 40 and 20$\%$ of the maximum photometry obtained in an 8-pixel aperture centered on the star. Thus, applying those filters will results in the removal of all frames for which the aperture photometry in an 8-pixel aperture centered on the star is above the threshold. For this particular case, the best filtering levels are reached at 60$\%$ for mean background subtraction and 80$\%$ for PCA background subtraction.

We can see in the top panel of Figure \ref{CC} that PCA background subtraction provides the best results for both filtered and non-filtered curves. For the optimally filtered case, the improvement obtained with PCA can go up to 1.2 while for the unfiltered case, the improvement can go up to 1.7. It is however important to remember that these contrast curves are contrast limited for the whole range of angular separation, due to the small field of view of our datasets. Thus, higher improvement factors are expected on background-limited regions.

As for the ADI case, we compared the contrast curves obtained through full-frame PCA PSF subtraction with RDI. For this analysis, we used the dataset of calibrator HD108381, which is composed of 8000 frames of 60 ms exposure time. In the RDI case, both the scientific and calibrator datasets are background subtracted in the exact same way. We thus match the calibrator observations which have been background subtracted with the same pre-subtraction, the same PC number for background subtraction and the same filtering level to the corresponding scientific dataset.  We present the results for this section in the middle panel of Figure \ref{CC}. In the case of PCA background subtraction, the best results were obtained with 10 principal components. The number of principal components for PSF subtraction are respectively 8 and 10 for the PCA-background-subtracted cube, and 6 and 10 for the mean-background-subtracted cube.

\begin{figure*}[htbp]
    \centering
    \includegraphics[width=1\linewidth, trim = 0 90 0 100 ]{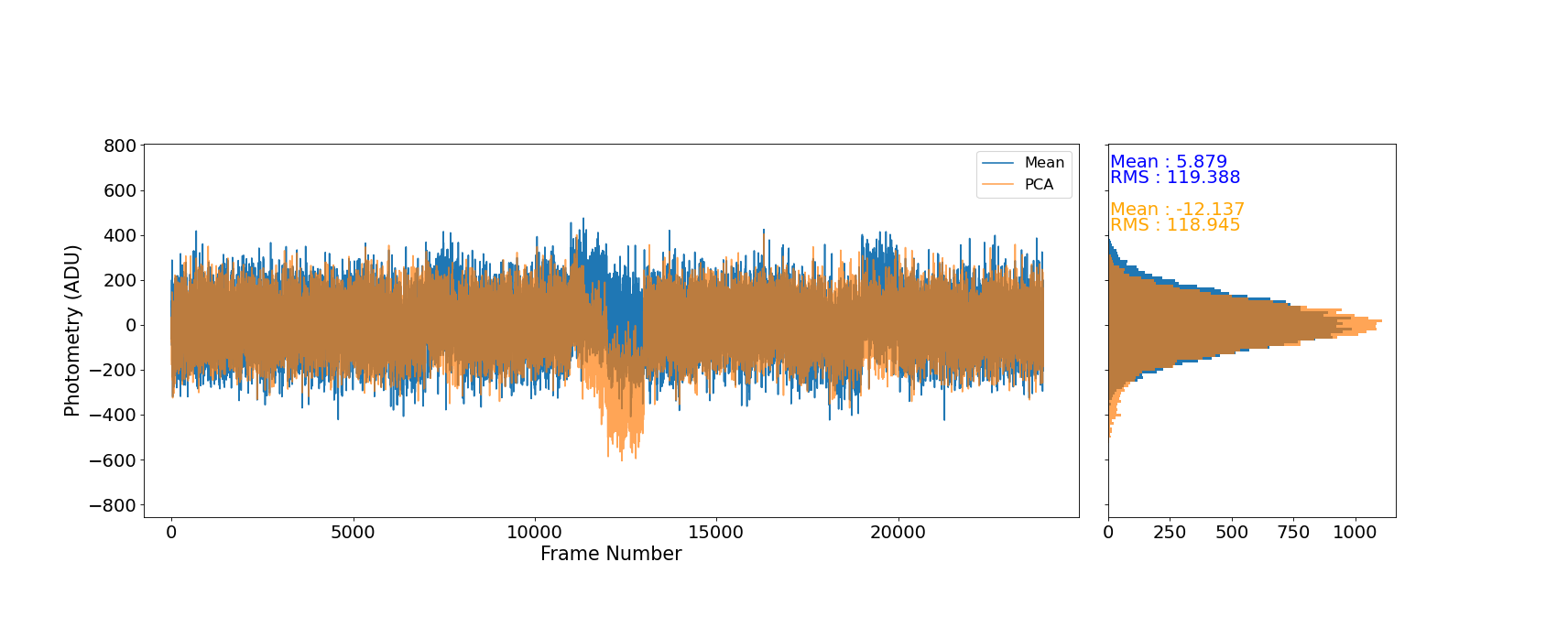}
    
    \includegraphics[width=1\linewidth, trim = 0 90 0 100]{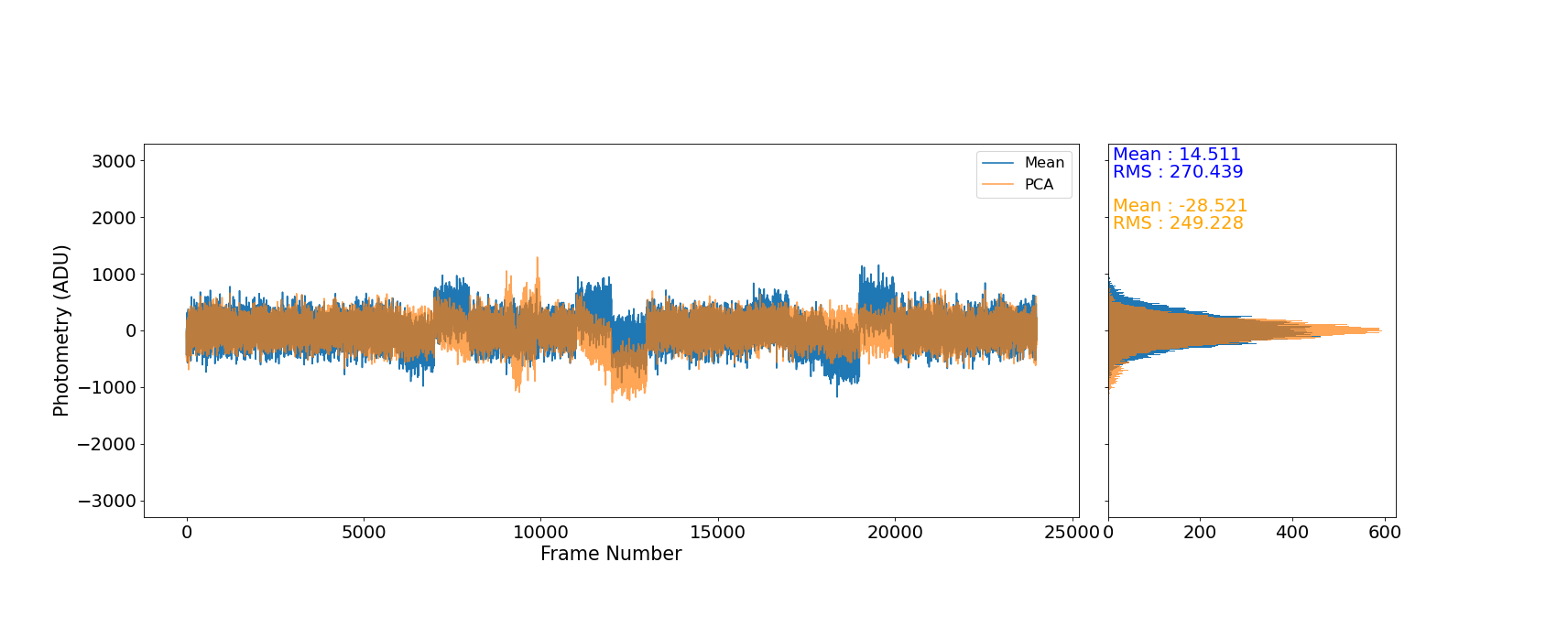}
    
    \includegraphics[width=1\linewidth, trim = 0 90 0 100]{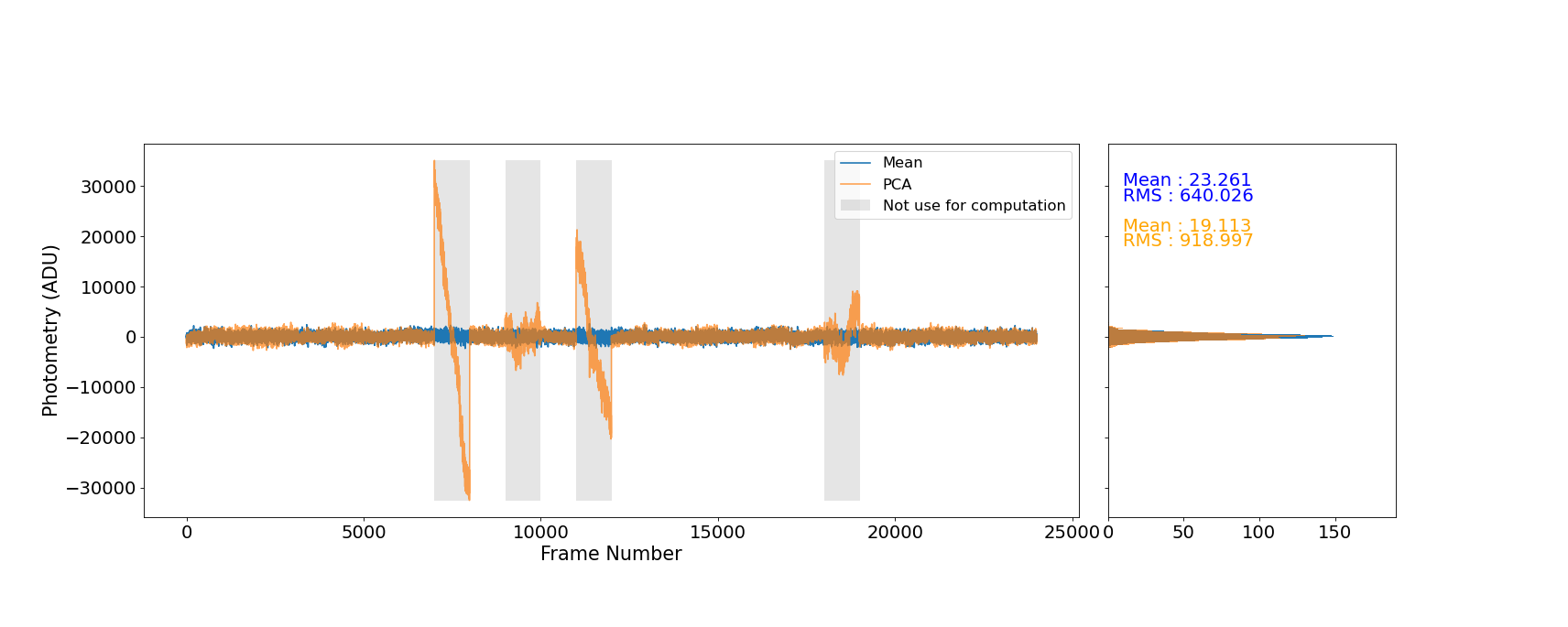}
    \caption{Aperture photometry per frame with apertures of 8 (top panel), 13 (middle panel) and 32 (bottom panel) pixels in radius for Mean (blue) and PCA (orange) background subtraction with a mask of 9, 13 and 32 pixels in radius, respectively. The grey boxes show the groups considered as outliers which are not used for mean retrieval and RMS for both mean and PCA background subtraction. The calibration factor for sensitivity is 1 ADU for 0.3 mJy.}
    \label{VisComp9}
\end{figure*}

We can see in the middle panel of Figure \ref{CC}, that, as in the ADI case, PCA background subtraction can bring an improvement to the contrast curves obtained with mean background subtraction. In the non-filtered case, we obtained an improvement factor up to 1.2. However, we obtained improvement factors from 1.1 at large separation to 1.3 at small separation between the two optimally filtered curves. As in the ADI case, these contrast curves are contrast limited for the whole range of angular separations.

For the analysis without PSF subtraction, we only de-rotated the frames, and form the final image with the median frame of the de-rotated dataset. As there is no self-subtraction for this method, the throughput is everywhere at 1. The main benefit of this method is to analyze the effect of the background subtraction without the biases of the PSF subtraction. We thus present the results of this method in the bottom panel of Figure \ref{CC}. For the filtered (dashed) curves, a filtering of 20$\%$ for the mean background subtraction and for the PCA background subtraction has been applied. 

Without PSF subtraction, since the residual starlight is not removed, the contrast strongly dominates. This can be seen at close separation, where the contrast is the most problematic, and PCA background subtraction does not improve the contrast curves. On the other hand, at larger separation where the impact of the contrast diminishes, PCA background subtraction improves the contrast curves, as expected.

\subsection{Aperture photometry analysis}\label{sec:APPhot}

In this section, we display the results of our comparison of Mean and PCA background subtracted datasets with aperture photometry on the empty dataset presented in Section 2.2. All apertures are centered on the center of the frame, where the star would be located if any were present.

\subsubsection{General photometry}\label{sec:Ap General}

We first discuss the case of general photometry, where the photometric aperture, background annulus (for classical, mean background subtraction) and mask (for PCA background subtraction) are optimized for the size of the astronomic source observed (e.g., star, galaxy). The mask has therefore the size of the source, and the background annulus inner radius is set to be exactly one pixel larger than the source extension. The outer radius of the background annulus is computed to match as closely as possible the mask and the annulus areas (no fractional pixels). For this case, we studied different possible source sizes, ranging from 8 pixels to 32 pixels in radius. In the case of mean background subtraction the photometry is computed as the difference between the photometry obtained in the central aperture and the one obtained in the background annulus. Since their areas do not exactly match, we normalized those two quantities by the number of pixels contained in each. A pre-subtraction has been applied, before the PCA subtraction on both the scientific images and the background images. No mask and no pre-subtraction have been applied for Mean background subtraction. \\

Figure \ref{VisComp9} presents the results for three particular aperture sizes (8-pixel: optimized for point sources, 13-pixel: marginally extended sources and 32-pixel: extended sources).Background residuals structures appear in the curves and increase in magnitude for larger apertures. The RMS of our measurements are similarly degrading. It is in particular visible with the large spikes in the PCA curve for the largest aperture. In the case of mean background subtraction, the background annulus effectively flattens the spikes and we obtain instead whole groups offset from zero. These structures in the photometry result from background structure residuals, and their variation, after the background subtraction is performed.

We now consider the mean retrieval of a source flux from the data. If no bias were present, the value should be consistent with zero within the statistical error $\sigma$ of the data, $\sigma = RMS/\sqrt{N_\text{f}}$, with $N_\text{f}$ the number of frames in the group. This value is, respectively, from the smallest to the largest aperture, 0.77, 1.75 and 4.13 for the mean background subtraction, and 0.77, 1.61 and 5.93 for the PCA background subtraction. This is not the case for any of our mean retrieval values, indicating that there is a bias present for both the mean and PCA background subtraction. It is however interesting to note that in the case of PCA, for the 8-pixels and 13-pixels apertures, if we exclude the easily identified outlier groups , the mean retrieval then respectively become 0.52 and 0.48, consistent with zero within the statistical error.\\

\begin{figure}[t]
    \centering
    \includegraphics[width=1\linewidth, trim = 20 0 0 0]{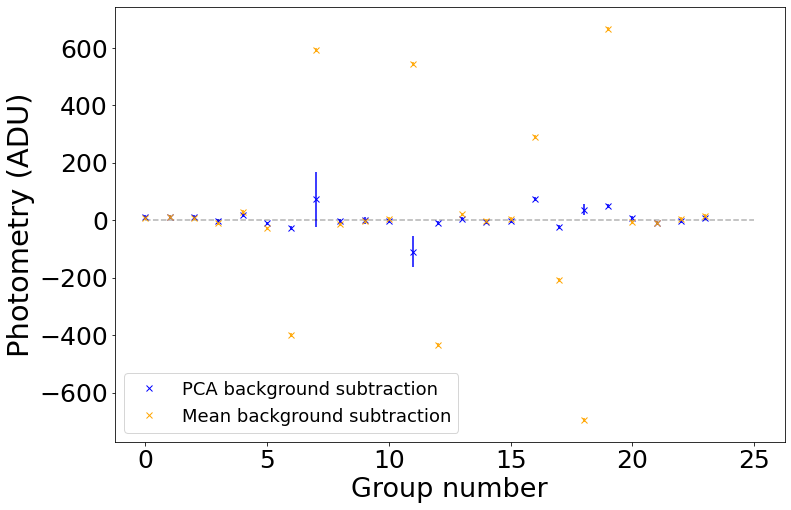}
    \caption{Comparison of the mean retrieval ADU values per group of 1000 images and their respective errors bars with an aperture of 13 pixels in radius, for mean background subtraction (orange points) and PCA background subtraction with a mask of 32 pixels in radius (blue points).}
    \label{VisuNod}
\end{figure}

It is important to note that each of these long sequences, and their mean retrieval value, are only one random realization of the photometry and is thus affected by the quasi-statistical background bias and by large, random errors. As the quasi-random effect of the background bias may result in a positive or negative offset with the statistical mode of its probability distribution at zero, a single measurement close to zero may be serendipitous and is not proof of a lack of background bias. As a consequence, they are not suitable for comparing the performance of the methods as a worse performing method can get “lucky” and produce a more accurate value than a generally better performing method due to random errors. It is thus not surprising that PCA produces a mean retrieval value that is further from zero than the result from mean background subtraction in two of the three cases, and not a reliable indication that PCA performs worse than mean background subtraction. To evaluate more reliably the performance of the two methods, we perform a statistical analysis on a per-group basis in Figure \ref{VisuNod}.\\

\begin{figure*}[t]
    \centering
    \includegraphics[width=0.52\linewidth]{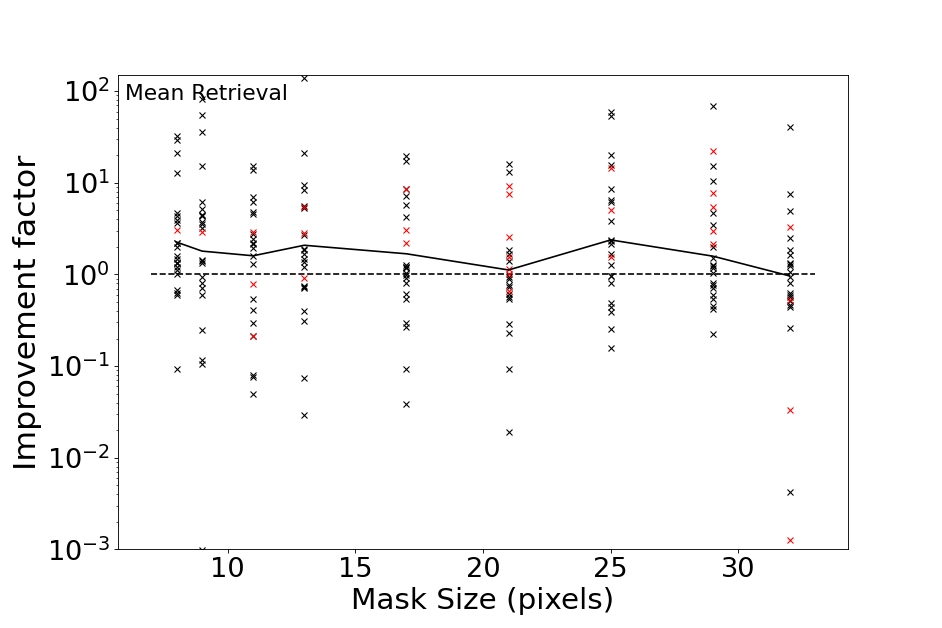}\includegraphics[width=0.52\linewidth]{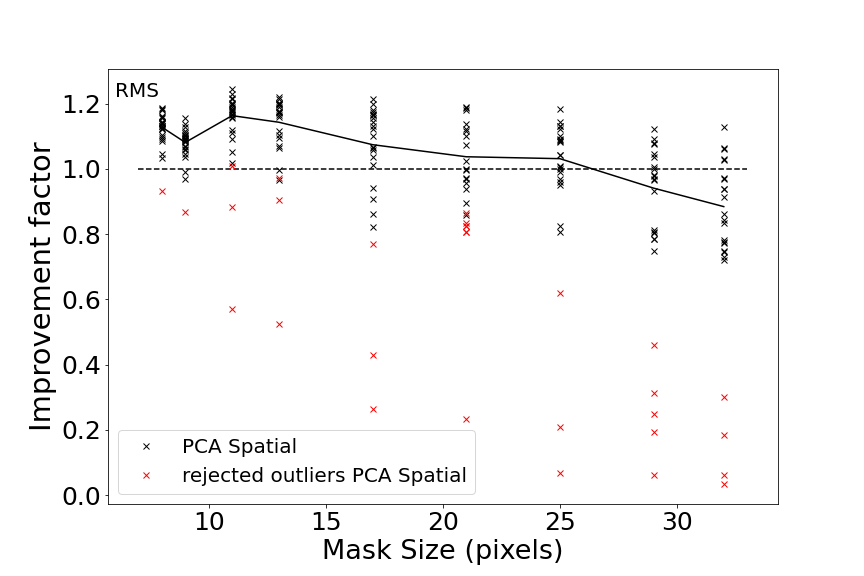}
    \caption{Improvements factors when using PCA background subtraction instead of mean background subtraction for mean retrieval (left panel) and RMS (right panel). Each point represents the ratio of the value obtained for PCA background subtraction and mean background subtraction. The background annulus inner radius is always one pixel larger than the mask radius. For each dataset, with its given mask and annulus size, we rejected up to 5 outliers which are here represented in red. These outliers are determined with respect to their RMS. The solid line represents the geometric mean of the improvement factors of individual nods, without the outliers.}
    \label{Imp Fact G}
\end{figure*}

In order to better compare the two methods, we computed, for each group of a thousand image, the mean retrieval value and its RMS. This comparison allows both to estimate the bias remaining in the data, if the points are significantly offset off zero, and the uncertainty of those values. Those results are shown in Figure \ref{VisuNod}. When using a 8-pixel or a 13-pixel aperture, some groups of images, when background subtracted with the mean method, are significantly offseted from zero. An even more problematic characteristic is that the error bars of those measurements failed to account for their dispersion. On the other hand, with the PCA background subtraction, those groups are much closer to zero.

Figure \ref{VisuNod} shows a clear improvement when using the PCA background subtraction instead of the classical mean background subtraction. Indeed the uncertainty of the HOSTS survey is dominated by the scattering of the measurement rather than by the measurement uncertainty themselves. It is to be noted however, that Figure \ref{VisuNod} present the results for the 13-pixel aperture only. The 8-pixel aperture present very similar results but for the scale at which the different measurements are scattered. For the 32-pixel aperture, on the other hand, the difference between mean and PCA background subtraction became less significant, in particular due to one extreme outlier. However, it is difficult with such plots to quantitatively estimate the improvement which is obtained with PCA background subtraction. We thus decided to compare instead the ratio between the values obtained with the mean background subtraction and the PCA background subtraction so to obtain improvement factors. We present those results in Figure \ref{Imp Fact G}

We compute improvement factors of the mean per group as the ratio between the absolute value of the mean from the classical background subtraction and the absolute value of the mean from the PCA background subtraction.  For the RMS, we compute the improvement factor as the ratio between the RMS from the classical background subtraction and the RMS from the PCA background subtraction. The measurements from the individual groups are still randomly distributed around zero and the individual improvement factors thus distributed between zero and infinity with a mode below one (for a degradation) or above one (for an improvement).  The results are thus still affected by large statistical noise.  We then compute the geometric mean of the improvement factors over all groups to suppress the statistical noise. We present those results in Figure \ref{Imp Fact G} for the mean retrieval (left panel) and the RMS (right panel) We rejected the outliers with the higher RMS as could be done on a similar analysis of actual source photometry on a science target. However, to prevent rejecting too much exposure time, we limit this number of outliers to 5 (20$\%$ of data). 

As we can see on the left panel of Figure \ref{Imp Fact G}, the mean retrieval benefits from a PCA background subtraction. We obtain on average, an improvement factor between 1.3 and 2.4 for the mean retrieval. We do not see any clear tendency on the range of mask sizes except for the largest one, for which PCA background subtraction provides results about the same quality of a standard mean background subtraction. On the right panel of Figure \ref{Imp Fact G}, we can see that the RMS is much less impacted by PCA background subtraction than the mean retrieval. This is because the RMS of the individual photometry measurements per image is dominated by background photon (Poisson) noise. PCA is powerful in removing subtle biases in the data that become significant when averaging a large number of frames or integrating for a long time, but have little effect on the RMS of the photometry of individual frames that is dominated by the individual frames' noise. However, the smallest apertures and masks, which accumulate the least Poisson noise, still benefits even if slightly from PCA background subtraction. For the largest masks however, we observe a slight degradation. These two plots shows that even for a reasonably large aperture and mask, PCA can provide significant improvements, and that its effectiveness decreases only for large masks compared to the total size of the frame. Thus, in the case of a not too extended emission, PCA background subtraction is much more effective than mean background subtraction for aperture photometry.

\begin{figure*}[t]
    \centering
    \includegraphics[width=0.52\linewidth]{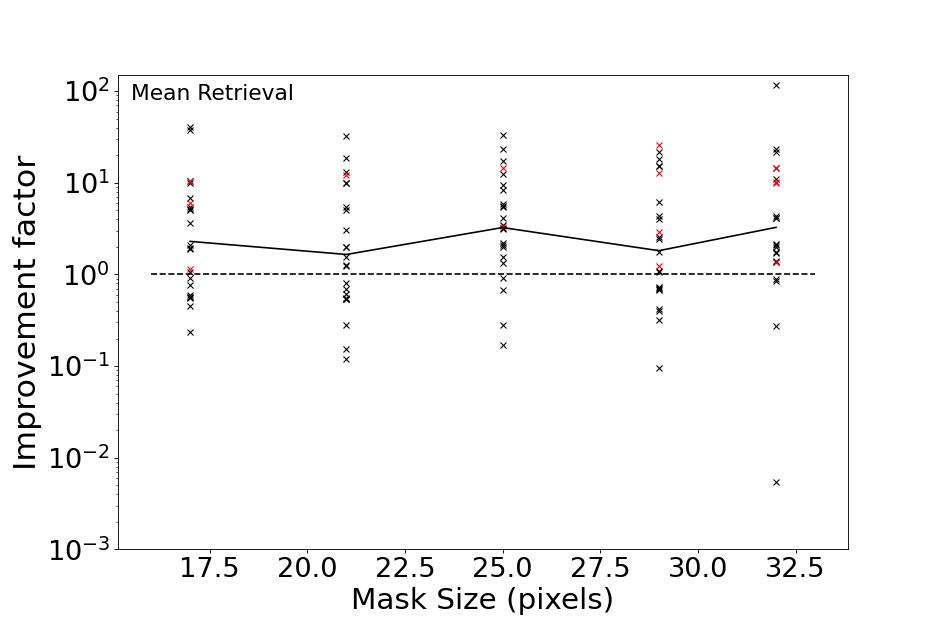}\includegraphics[width=0.52\linewidth]{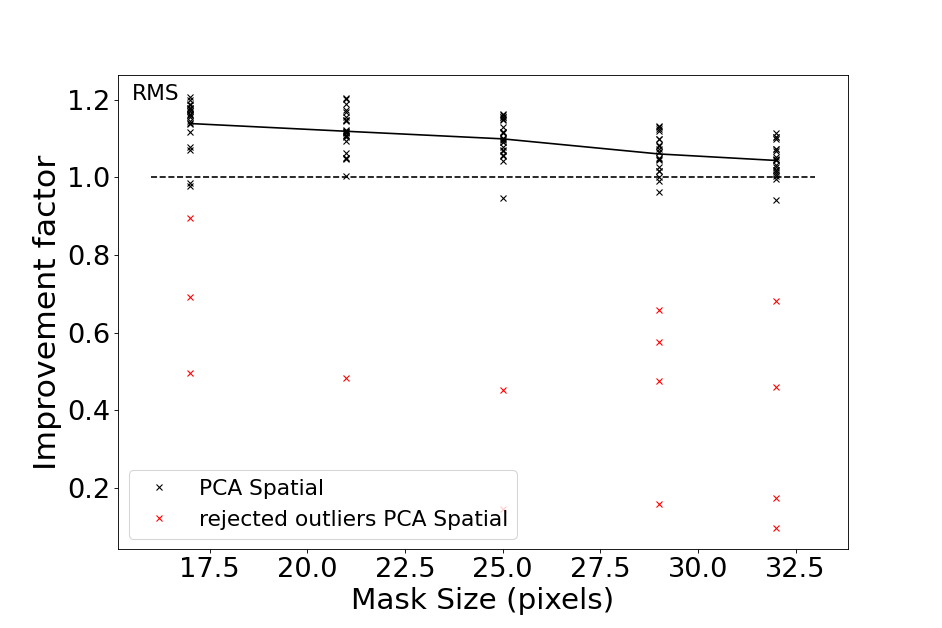}
    \caption{Same as Figure \ref{Imp Fact G}, but with a fixed aperture size (8pix) and varying size of the PCA mask and background annulus.}
    \label{Imp Fact H8}
\end{figure*}

\begin{figure*}[t]
    \centering
    \includegraphics[width=0.95\linewidth, trim = 0 90 0 100]{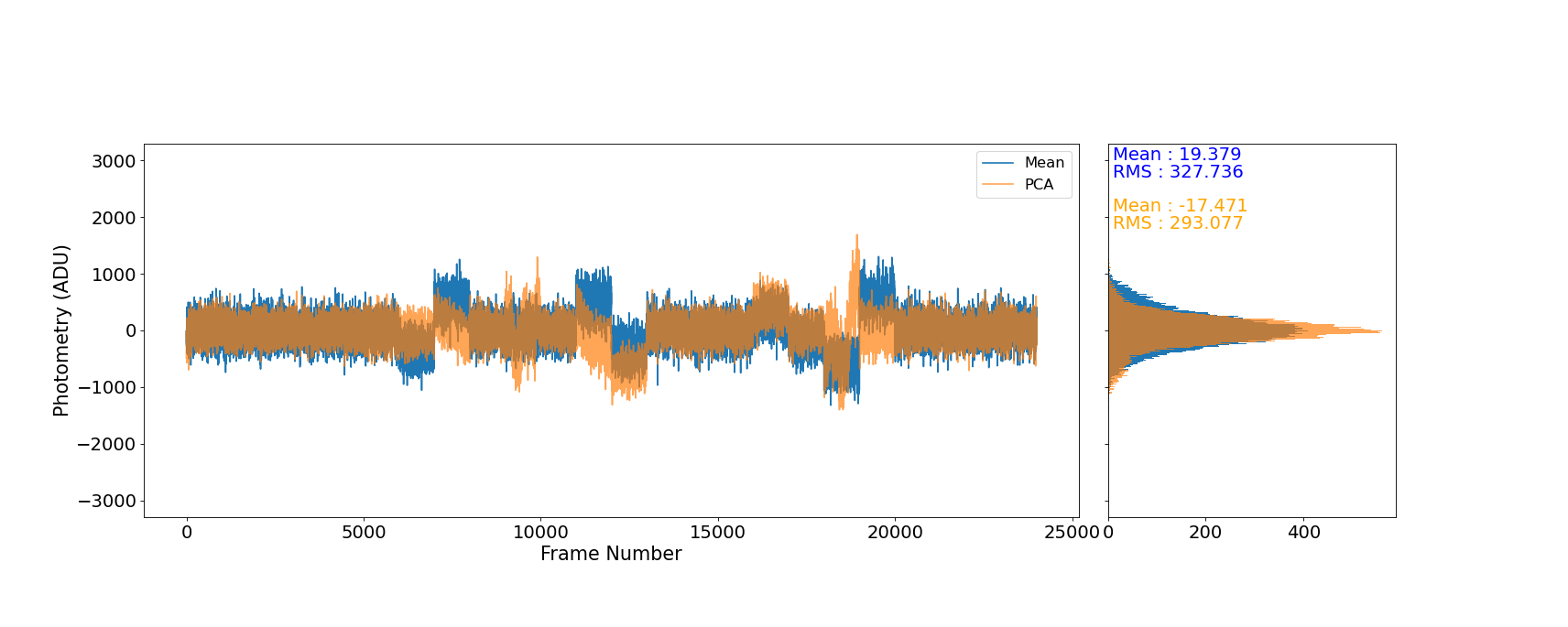}
    \includegraphics[width=0.95\linewidth, trim = 0 90 0 100]{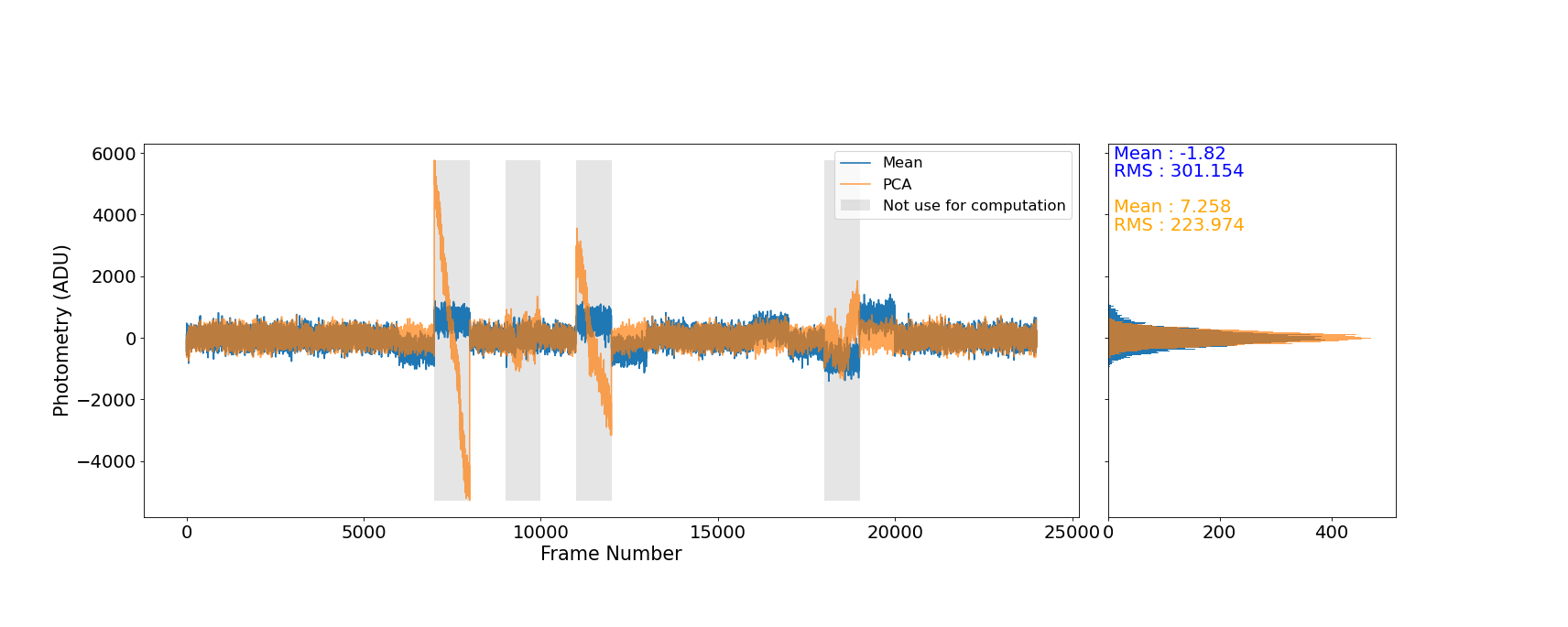}
    \caption{Similar as Figure \ref{VisComp9} but with a fixed aperture size of 13 pixels  in radius and mask radii of 17 (top panel) and 32 pixels (bottom panel). The calibration factor for sensitivity is 1 ADU for 0.3 mJy.}. 
    \label{VisComp13}
\end{figure*}

\subsubsection{Varying mask sizes for a fixed aperture size}\label{sec:Ap HOSTS}

Here, we explore the impact of varying the mask size for a given, fixed aperture size.  The mask is a significant constraining factor for the PCA performance, limiting the amount of information available for the PCA background subtraction.  Scientifically, this is also an important study for the specific case of LBTI nulling interferometry for the HOSTS survey as described in  Section \ref{sec:appli_ApPhot}, where the mask needs to cover all plausible disk emission while a smaller aperture can be used to optimize the expected signal-to-noise or constrain the emission within a certain radius.  More generally, one may want to adjust the mask size to the specific science case and data properties as one would do with the background annulus for classical background subtraction.  This impact of varying mask size thus needs to be understood.

As for the general case, we first compare the photometry across the whole sequence, and then compute improvement factors based on a group-by-group analysis. In Figure \ref{Imp Fact H8} we present the latter results for the 8-pixel photometric aperture and a range of mask and background annulus sizes optimized for the conservatives apertures found for the HOSTS targets. 

The left panel of Figure \ref{Imp Fact H8}, shows that PCA background subtraction provides significantly better results than the mean background subtraction with about a factor 1.7 to 2.5 improvement over the range of mask sizes. These improvement factors are mostly valid for the whole range of aperture, mask and annulus sizes we probed. Even for the largest aperture we do not observe any degradation on the mean retrieval. On the right panel of Figure \ref{Imp Fact H8}, we present the results for the RMS. We can see on this panel that PCA tends to improve the RMS for all aperture sizes. However, the improvements are quite small and one can consider that the PCA method does not significantly impact the RMS compared to the mean method.The 13-pixel aperture being the optimal one for the HOSTS survey \citep{ertel_hosts_2020}, we here present the detailed analysis for this aperture size together with two other mask sizes: 17 and 32 pixels, respectively the smallest and largest conservative apertures in the sample of the HOSTS survey. 

\begin{figure*}[t]
    \centering
    \includegraphics[width=0.50\linewidth]{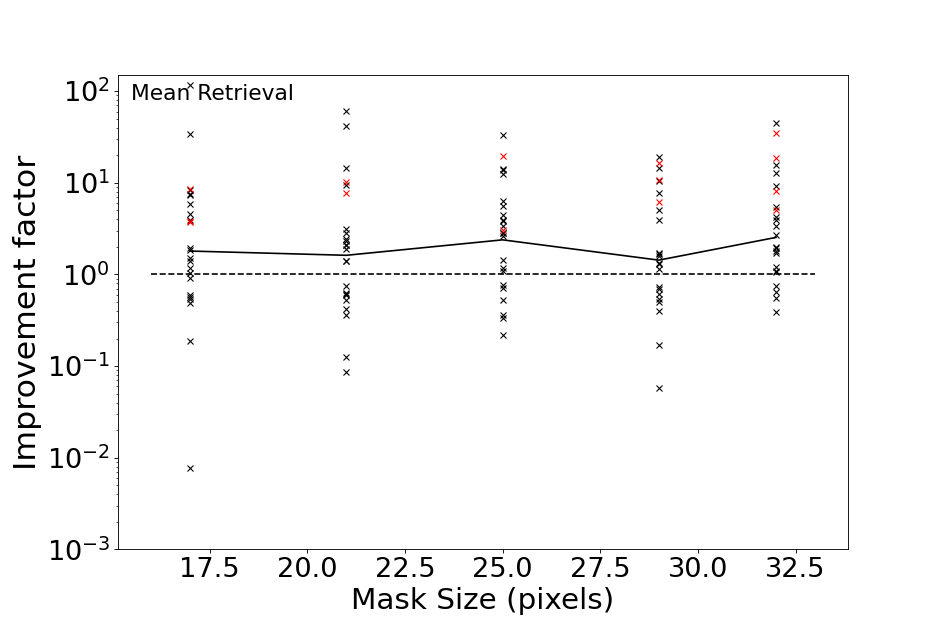}\includegraphics[width=0.50\linewidth]{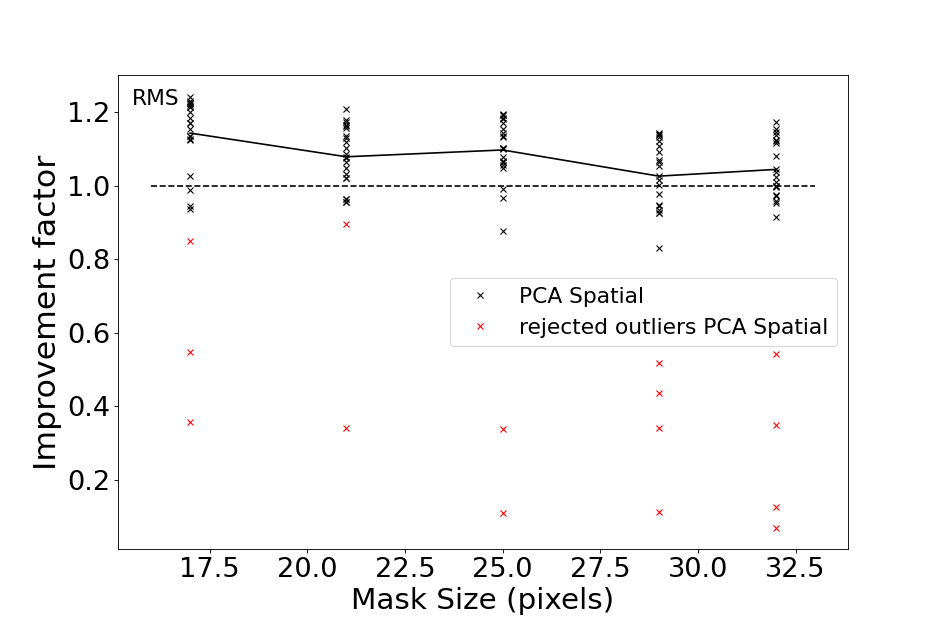}
    \caption{Similar as Figure \ref{Imp Fact H8} for a 13-pixel aperture.}
    \label{Imp Fact H13}
\end{figure*}

The top panel of Figure \ref{VisComp13}, shows a certain number of structures in both curves, the ones for mean background subtraction being flatter due to the background annulus correction. As previously explained, we discarded the worst groups in terms of RMS for the computation of the mean retrieval and the RMS on the whole dataset. Outside of the group impacted by those structures, we can see however than PCA provides a more stable photometry. The general mean retrieval on the first panel is only slightly better for PCA than with the mean method. For the second panel, it is even slightly degraded. This is due to the groups averaging themselves out better in the case with the mean background subtraction than in the case of the PCA background subtraction. However, this does not translate the real quality of the mean retrieval in the dataset. This is why we prefer the use of the group per group analysis which provides more measurements and thus a more precise estimate of the mean retrieval quality and of the improvement brought by the PCA background subtraction. The group per group analysis over the range of conservative apertures is shown in Figure \ref{Imp Fact H13}. 

\begin{figure*}[t]
    \centering
    \includegraphics[width=0.50\linewidth]{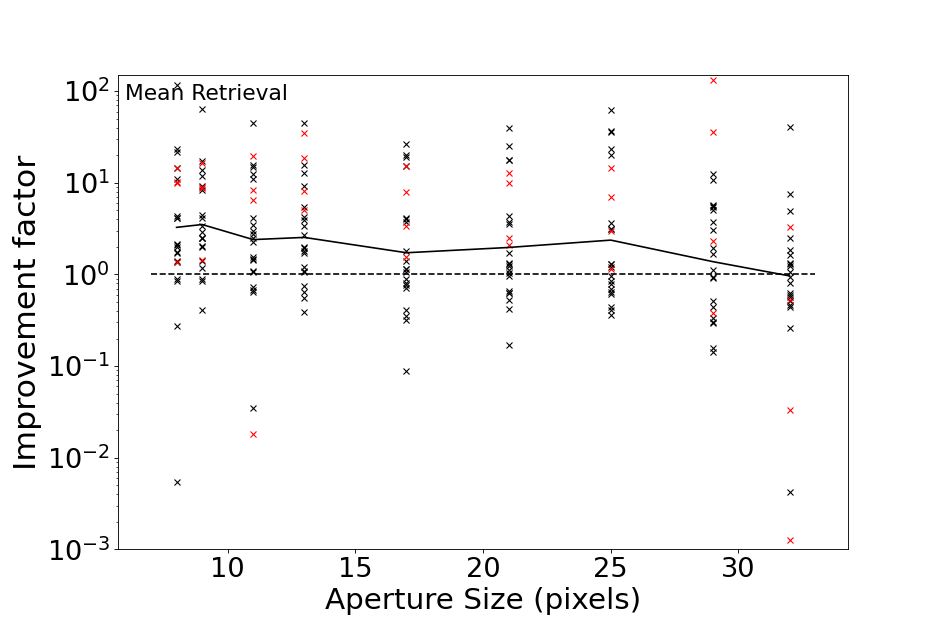}\includegraphics[width=0.50\linewidth]{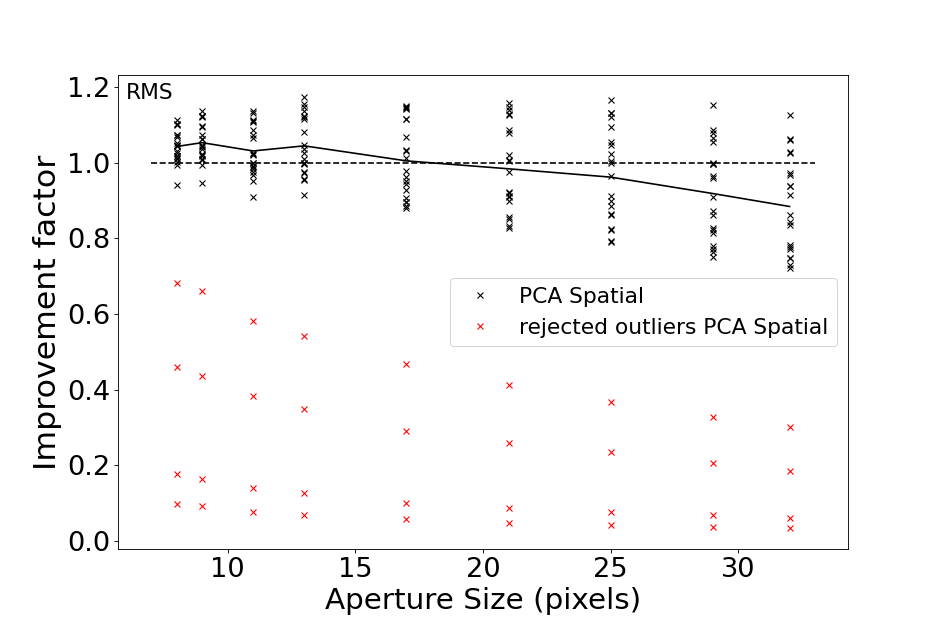}
    \caption{Similar as Figure \ref{Imp Fact G} but with a fixed mask radius of 32 pixels and varying aperture sizes.}
    \label{Imp Fact B}
\end{figure*}

The left panel of Figure \ref{Imp Fact H13},shows, as in Figure \ref{Imp Fact H8}, that PCA is more effective than mean background subtraction for the mean retrieval. Here, we obtain about a factor 2 improvement over most of the range. On the right panel, for the RMS, we can see that PCA still performs slightly better than mean background subtraction, with similar factors to the 8-pixel case. 

The case of the aperture of 32 pixels in radius is similar to the general case, already shown in Section \ref{sec:Ap General} for which we have demonstrate that the mean retrieval remains almost the same as in the mean background subtraction case and the RMS is only slightly degraded. This largest aperture put aside, in this second case too, PCA background subtraction provides significantly better results with improvement factors ranging from 1.7 to 2.5 for most apertures and mask sizes.

\subsubsection{Varying aperture size for a fixed mask size}\label{sec:Ap BLeo}

It is also interesting to investigate the effect of a fixed, large mask size and varying size of the photometric apertures.  This can be used to determine the radial distribution of extended emission (e.g., exozodiacal dust in HOSTS data, extended galaxies), or for curve-of-growth photometry of point sources.  Here we use the case of $\beta$\,Leo from the HOSTS survey as a proxy and present the results for the conservative aperture of this star. The mask radius is thus set to 32 pixels and the inner radius of the background annulus to 33 pixels. These results are presented in Figure \ref{Imp Fact B} . 

An interesting feature can be seen on the left panel of Figure \ref{Imp Fact B}. As we can see, with a fixed mask size, we still observe a degradation of the improvement factors toward large aperture. The PCA background subtraction is thus more sensitive to the increase of the aperture size than the mean background subtraction. In terms of improvement factors we obtain a significant improvement over most of the range of aperture sizes. If we exclude the last aperture size (32 pixels in radius) we obtain improvement factors ranging from 1.4 to 3.2. This case shows that even with a really large mask the small apertures benefit strongly from the PCA background subtraction. Additionally we do not observe any degradation when using the PCA background subtraction instead of the mean background subtraction, over the whole range of aperture sizes in this case. 

As we can see on the right panel of Figure \ref{Imp Fact B}, the RMS does not significantly benefits from PCA background subtraction with such a large mask. However, we do not observe any significant degradation either. 

\subsection{Ring apertures}

In a recent effort to improve the null measurement of the HOSTS survey, we have begun to use annulus photometry instead of using filled, circular apertures. The advantage of this is that the ring-like photometric aperture can be placed at the region where the extended dust emission is expected, while ignoring a significant fraction of the residual star light. Furthermore, the use of different ring radii may allow one to characterize the radial distribution of the dust for the more extended systems (Faramaz et al, in prep.). In order to better estimate the real improvement brought by PCA with this new geometry for null measurement, we also performed annulus photometry instead of circular aperture photometry. We obtained similar improvement factors for both mean retrieval and RMS. However, for larger apertures, we observed slightly better results. In particular, we do not observe any degradation when using the annulus apertures instead of circular apertures. Thus we expect the PCA background subtraction to improve the results of the null measurements with annular aperture at least as much as it would for circular aperture.

%% file: Sections/discussion.tex
\section{Discussion}\label{sec:discu}

\subsection{Limitations and path forward}\label{sec:limitpath}

Figures \ref{VisComp9} and \ref{VisComp13} have shown that PCA background subtraction can fail for particular groups of images with large spikes in the frame per frame photometry. We investigated potential sources for these failures. We have shown that the groups with the strongest spikes can be easily identified by their increased RMS of the photometry. This provides an option to deal with them by simply ignoring the data in those spikes. A solution that would retain all data would however be preferable. 

\begin{figure*}[t]
    \centering
    \includegraphics[width=1\linewidth]{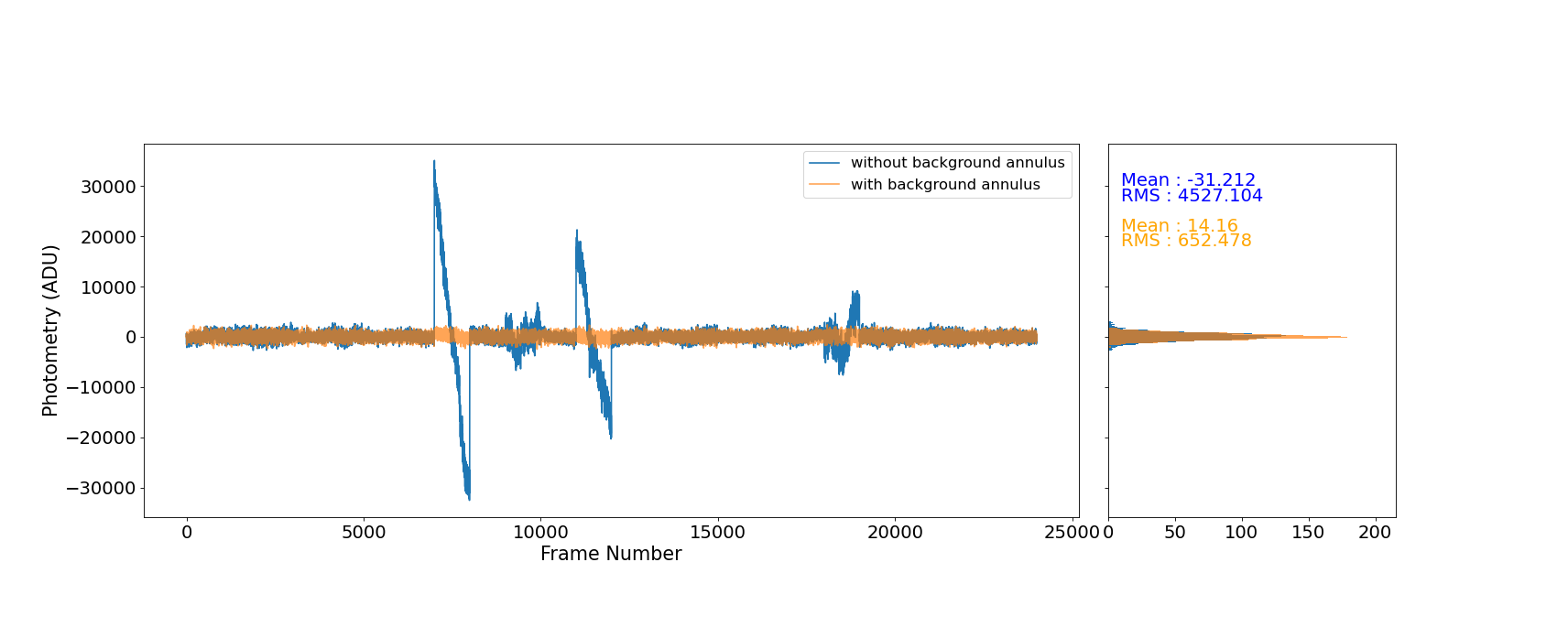}
    \caption{Similar as Figure 1 but comparing PCA background subtraction with (orange curve) and without (blue curve) background annulus. The aperture and mask size are 32 pixels in radius and the inner radius for the background annulus 33 pixels. The calibration factor for sensitivity is 1 ADU for 0.3 mJy. }
    \label{BgAnn}
\end{figure*}

To investigate the origin of these failures, we applied first a background annulus to the aperture photometry after PCA background subtraction. This successfully suppresses the spikes similar to the case of classical, mean background subtraction. This is shown in Figure \ref{BgAnn}. We conclude that the spikes are not introduced by PCA but rather not as well suppressed as by the background annulus. We then explored different sources of the spikes in the images: 

\begin{enumerate}
    \item a compact, bright structure in the aperture, and under the mask which could thus not be corrected by PCA.
    \item a compact, bright structure located outside of the aperture and brought in the photometry of the aperture due to the correction.
    \item a largely extended structure, both outside and in the aperture, the brightness of which rapidly varies.
    \item a rapid overall brightness variation of the background.
\end{enumerate}

From those hypotheses we started to rule out the different options, in particular those with a compact structure.
Since the large spikes appear both for PCA and mean background subtraction when not using a background annulus, (2) is very unlikely. A mean background subtraction would not bring the photometry of this small bright structure into the photometry of the aperture. Thus those spikes would not be observed in the photometry with the mean method when no background annulus is used. 

In addition, the use of the background annulus for both mean and PCA background subtraction flattens the spikes which then only appear as small offsets. This behavior also rules out (1) since a small, bright spikes in the photometric aperture would not be flattened by an annulus of background which does not contain this structure.

In addition we tried to change the location of the aperture in the image. With this change of position some spikes appear for different groups and some present with the centered aperture disappeared. However, for the group  presenting the strongest spike (from frame 10000 to 11000) in Figure \ref{VisComp9} and \ref{VisComp13}, the spike remains at whichever location we placed the photometric aperture. As a small bright structure would not appear in all the apertures at different location on the image, this rules out the possibilities (1) and (2) of a small structure.

We thus conclude that the reason for those failures must be of type (3) or (4), with an extended emission the brightness variation of which is too significant or too fast for PCA to correct it. We also found that, even for the smallest aperture (8 pixels in radius), this structure appears when using a larger mask.\\

The simplest solution to overcome this problem would thus be to limit the size of the mask but this would be a very strong constraint on possible ways to apply the method. Another possibility is to use a background annulus with the aperture photometry in addition to the PCA background subtraction. This however defeats the purpose of PCA as the background annulus adds photon noise and at least some of the background bias back into the data.  The results shown in Figure \ref{BgAnn} use an optimal configuration in which the inner radius of the background annulus is one pixel larger than the photometric aperture radius. In cases such as discussed in Section \ref{sec:Ap BLeo} in which the mask and the inner radius of the background annulus are larger than the photometric aperture, this technique does not work effectively and the spikes remain.

The optimal solution would be to address the rapid brightness variation of the overall image or of the extended structure over time. For this purpose, we will develop a temporal-PCA approach which would reduce significantly background brightness variation over time. This temporal PCA would thus build its principal components on the variation of individual pixels through time. Since the large-scale variations of the background correlate for a large number of pixels that are located close to each other in the image, the temporal PCA will identify these as dominant features in the data that can then be removed. We also expect this procedure to reduce the differences between the background exposures and the sources exposures (as the variation of pixels over time will be reduce for both) thus allowing for a better correction than with the spatial PCA alone. 

Since the actual source flux variation in the image is not used afterward in the case of regular photometry, it may not have to be masked during the temporal PCA step. For nulling interferometry, the instrumental null depth does, however, vary over time and this variation is used for the statistical analysis of the data (null self calibration,\citealt{defrere_2016}). While PCA may potentially provide an alternative way to analyse the null variation that could be explored, masking the star will still be required if the null variation is to be preserved for further analysis. The dust emission region may, however not need to be masked similarly to the case of regular photometry. This provides an opportunity to eliminate the mask, or at least reduce its size, which is a major performance limitation of the PCA method.

As a first estimate of the possible improvement, we used the background annulus already used for mean background subtraction. This is shown in Figure \ref{BgAnn}. We can see on this figure that the spikes disappear and the RMS is reduced by about a factor 3. This estimate shows how much we can improve our results by completely removing the spikes. It is possible that the temporal PCA correction performs slightly worse, with small photometric variations left which would degrade the RMS, but still allow to use those groups of images and improve the mean retrieval. On the other hand, the temporal PCA correction should not only affect the groups of images with the spikes in photometry, but also reduce the RMS in the less extreme groups of images.  On a group per group analysis to estimate performances with and without a background annulus, we observe that the mean retrieval is significantly degraded on most of the groups with the use of the annulus. However, we do not expect such effect with the temporal PCA correction. Indeed the bias comes from the spatial difference between the aperture location and the background annulus location. With the use of the temporal PCA correction, only the principal components would be computed at a different location while the projection will be performed on the aperture location, thus limiting this effect.

As shown in the case of aperture photometry, some structures are not perfectly removed by the spatial PCA background subtraction. These likely large-scale, time-variable structures causing the spikes on the photometry sequence will also affect the HCI case, where they may be limit to the achieved contrast or possibly result in spurious detection. Further investigation of the benefits of temporal PCA in HCI data sets is likely of merit as well.

\subsection{Application to HOSTS and nulling data}

The HOSTS survey was able to put an already much stronger constraint than previous missions on the median zodi level around nearby stars. This median level is of primary importance for future direct imaging missions for exoplanets since a large amount of zodiacal dust in a system can easily outshine and hide an earth-like planet. Even if the HOSTS survey has already demonstrated that the median zodi level would not definitively prevent direct-imaging mission to image earth-like planet, the sensitivity of its measurement would need to be improved by a factor 2 in order to address the feasibility of direct-imaging missions currently under development such as the Habitable Word Observatory (HWO).

This sensitivity improvement would translate in our analysis by an improvement of the mean flux retrieval. Thus, it is interesting to see that in all cases, except for the largest aperture (32 pixels in radius), the improvement factors on the mean retrieval are about 2. From the conservative apertures radii used in \cite{ertel_hosts_2020} only 6 stars over 38 have conservative apertures equal or above 32 pixels in radius. We thus expect this factor 2 to translate directly to nulling measurements over all three apertures for most stars. In future work, we intend to apply this new method to the whole HOSTS survey target sample and re-analyze the data with the new PCA background subtraction. From this factor 2 to 3 improvement on the mean retrieval, we expect to both put stronger constraints on the already detected exozodis, but also to increase the number of detections among the HOSTS target sample. Furthermore, we intend to perfect this method with the introduction of the temporal correction described in Section \ref{sec:limitpath}, in addition to the spatial correction described in this work. This new analysis of the HOSTS survey will strongly benefit future direct imaging missions. 

We also believe that developing similar PCA-based approaches for single-mode-fiber fed nulling instruments, such as the Nulling Observations of exoplaneTs and dusT instrument (NOTT, \cite{Defrere2018, Defrere_2022}), and missions such as the  \textit{Large Interferometer For Exoplanets} (LIFE, \cite{2022A&A...664A..21Q}), would be strongly beneficial to push their sensitivity and improve their science return by improving their detection limits and reducing the time needed for individual observations.

\subsection{General applications}

The results shown in this article are valid for the nulling-interferometric data of the HOSTS survey. However, this method can be applied to a much wider range of data type and observations. In this section, we discuss how the improvement obtained with PCA might change in function of different data parameters such as the predominance of background bias relative to its RMS in the data or observational parameters such as the nodding frequency or the time needed for an offset.

\begin{figure}[t]
    \centering
    \includegraphics[width=1\linewidth, trim = 0 0 0 0]{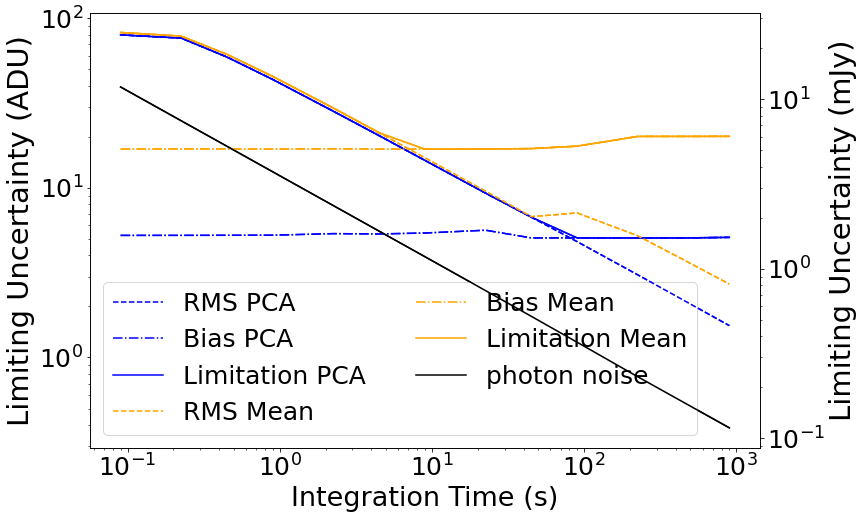}
    \caption{Main sensitivity limitation in background subtracted data, with an aperture of 13 pixels in radius for mean background subtraction (orange) and PCA background subtraction (blue). The photon noise is indicated in black.}
    \label{IntTime}
\end{figure}

The sensitivity is limited by several sources of noise. Here we estimate the contribution of different noise sources for both mean and PCA background subtraction. In particular we show in Figure~\ref{IntTime}, the effects of the photon noise, the ELFN noise (RMS) and the bias with respect to the integration time.

As we can see on Figure~\ref{IntTime}, the PCA background subtraction has very little effect on the RMS. However, it improves significantly the bias left after the background subtraction. With a sufficient integration time, this bias improvement directly translate in a sensitivity improvement. For both the datasets used in this article, the integration time is long enough so the sensitivity is limited by the bias for both mean and PCA. Similarly, all datasets from the HOSTS survey will be in the regime limited by the bias. Thus it is reasonable to expect similar improvement factors. With the temporal PCA additional correction we discussed in the previous subsection we expect to push the RMS closer to the photon noise limit. Such improvement would prevent being limited by the RMS for shorter integration times.  

\begin{figure}[t]
    \centering
    \includegraphics[width=0.9\linewidth, trim = 20 0 40 0]{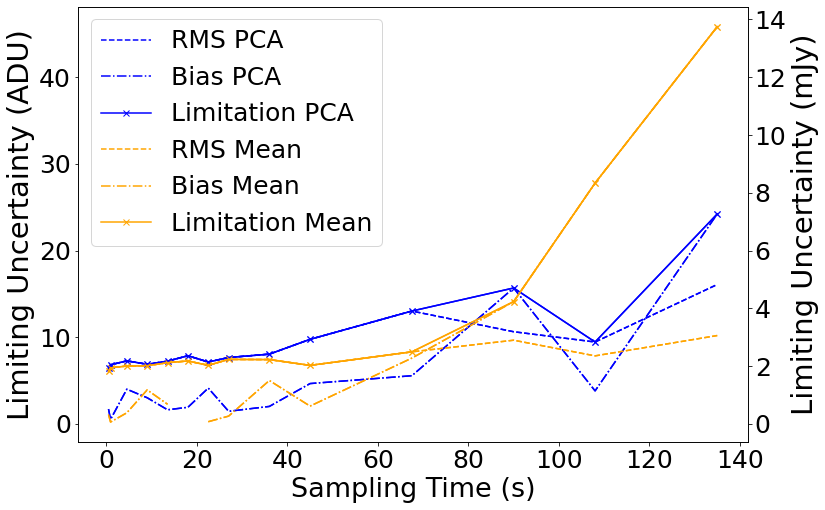}
    \caption{Sensitivity limit as a function of time for background subtracted data}
    \label{NodF}
\end{figure}

Another important parameter for the background subtraction is the nodding frequency. The shortest the frequency, the less the background will change in between background exposures and source exposures. However, in the case of PCA, having a larger library to build the principal components can be beneficial. In Figure~\ref{NodF} we show the effect of the nodding frequency on the limiting uncertainty. 

We can see in Figure~\ref{NodF}, that for short enough nodding frequency we remain within reasonable limits on the uncertainty. However, with longer periods, the uncertainty increase rapidly. An interesting result shown on this figure is that PCA background subtraction allow to use longer nodding frequency compared to mean background subtraction. The datasets used in this article have a nodding frequency of 45s. In this configuration both mean and PCA background subtraction remains in reasonable uncertainty limits. However, for the rest of the HOSTS survey the nodding frequency was about 90-120s. With this new configuration, we can see that using the PCA background subtraction becomes more advantageous. Thus, in general, using the PCA background subtraction would allow to use longer nodding periods. 

In the case of the background dataset, used for aperture photometry, we simulated groups of 1000 frames. However, since this dataset was taken with no sky offset, the simulated offset was instantaneous. To simulate a more realistic dataset, we thus tried different gap times during which a real offset would happen. We considered gap times from 0 to 140s for both mean and PCA background subtraction. As expected, we found that the quality of the background subtraction decrease with increasing gap time. However, this degradation is very similar for both background subtractions. We thus expect this parameter to have little to no impact on the improvement factors obtained with the PCA background subtraction. 

Finally the improvement achieved by PCA background subtraction would depend on the wavelength. \cite{hunziker} have applied a similar method in the $L$ and $M$ band, which benefits from a lower thermal background than in the $N$ band, and have found a smaller impact of the PCA background subtraction. Analogous studies should be perform with longer wavelength to determine the real impact of such method for longer waveband.

%% file: Sections/conclusion.tex
\section{Summary and conclusions}\label{sec:conclu}

This study shows that PCA thermal background subtraction can achieve significant improvement over mean background subtraction for both aperture photometry and high contrast imaging in the mid-infrared (N band). For the latter, we have demonstrated that a PCA background subtraction can improve the reachable contrast by a factor 1.2 to 1.7. We have shown that this improvement is significant for both very commonly used ADI and RDI PSF subtraction techniques but that without any PSF removal the contrast dominates too much to observe any significant improvement. \\

For aperture photometry, we have shown in particular that, without degrading the photometric precision, we can reach an improvement factor from 1.4 to 3.2 on the accuracy of the mean retrieval. Imperfect thermal-background subtraction has been shown to be a major sensitivity limitation of the HOSTS survey \citep{defrere_2016}. This limitation is mainly due to the bias on the individual, calibrated null measurements rather than their errors bars. With a factor 1.7 to 2.5 improvement on the accuracy of the mean retrieval over most of the range of apertures, we expect about a factor 2 improvement for those biases, and thus on the sensitivity. Further improvement of the nulling mode depends on the brightness of the star and will require further investigations. 

The approach presented in this work can be applied to a wide variety of existing data sets and future observations, since it only requires background observations that are regularly interleaved with the science observations (e.g., through nodding). All existing datasets with these characteristics would be suitable for applying this method. We strongly expect suitable existing datasets to benefit from this PCA background subtraction approach with similar improvement factors. Similarly, future datasets with these characteristics such as data taken by JWST, large ground-based telescopes and future ELTs, would also strongly benefit from this approach. An improvement factor up to 3 in sensitivity in thermal-infrared observations will make such observations up to 9 times less time consuming, and hence greatly improve the science return of these observatories.